\documentclass[11pt,a4paper]{article}
\pdfoutput=1
\usepackage{jcappub}
\usepackage{enumerate}
\def\lsim{\raise0.3ex\hbox{$\;<$\kern-0.75em\raise-1.1ex\hbox{$\sim\;$}}}
\def\gsim{\raise0.3ex\hbox{$\;>$\kern-0.75em\raise-1.1ex\hbox{$\sim\;$}}}

 \normalsize
\newcommand{\bmat}{\left(\begin{array}}
\newcommand{\emat}{\end{array}\right)}
\newcommand{\be}{\begin{equation}}
\newcommand{\ee}{\end{equation}}
\newcommand{\bea}{\begin{eqnarray}}
\newcommand{\eea}{\end{eqnarray}}

\usepackage{graphicx}
\usepackage{amsmath}
\usepackage{hyperref}
\usepackage{psfrag}
\usepackage{epsfig}
\usepackage{xcolor}
\usepackage{epsfig,amsmath}
\begin{document}

\title{A viable logarithmic \boldmath{$f(R)$} model for inflation }

\author[a]{M. Amin,}
\author[a]{S. Khalil}
\author[a,b]{and M. Salah}

\affiliation[a]{Center for Fundamental Physics, Zewail City of Science and Technology, 6 October City, Giza, Egypt}
\affiliation[b]{Department of Mathematics, Faculty of Science, Cairo University, Giza, Egypt}

\emailAdd{mamin@zewailcity.edu.eg}
\emailAdd{skhalil@zewailcity.edu.eg}
\emailAdd{mabdelmoneim@zewailcity.edu.eg}


\date{\today}

\abstract{Inflation in the framework of $f(R)$ modified gravity is revisited. We study the conditions that $f(R)$ should satisfy in order to lead to a viable inflationary model in the original form and in the Einstein frame. Based on these criteria we propose a new logarithmic model as a potential candidate for $f(R)$ theories aiming to describe inflation consistent with observations from \textit{Planck} satellite (2015). The model predicts scalar spectral index $0.9615<n_s<0.9693$ in agreement with observation and tensor-to-scalar ratio $r$ of order $10^{-3}$. Furthermore, we show that for a class of models, a natural coupling between inflation and a scalar boson is generated through the minimal coupling between gravity and matter fields and a reheating temperature less that $10^9$ GeV is obtained.}
\maketitle
\flushbottom
%%%%%%%%%%%%%%%%%%%%%%%%%
\section{Introduction}
%%%%%%%%%%%%%%%%%%%%%%%%%

Inflation is one of the most motivated scenarios for explaining the origin of structure formation in the Universe \cite{Brout:1977ix, Starobinsky:1980te, Kazanas:1980tx, Sato:1980yn, Guth:1980zm, Linde:1981mu, Albrecht:1982wi}. For more than three decades, the cosmological observations on the cosmic microwave background radiation and the large-scale structure in the universe have confirmed the predictions of inflation. The most common framework for inflation is based on a scalar field $\phi$ called inflaton \cite{Linde:1987aa, Liddle:2000cg} dominating the energy density of the universe and rolling down slowly along an almost flat potential $V(\phi)$. At the end of inflation the scalar field decays producing the known conditions for the standard hot Big Bang cosmology.  

This hypothetical scalar field remains  mysterious  and to construct an inflationary scenario, an extension of the standard model of particle physics is required, or alternatively, this scalar degree of freedom could be attributed to the gravitational sector. Brans-Dicke theory is probably the first well known theory of gravity that includes a scalar field. For a vanishing Brans-Dicke parameter, this theory is dynamically equivalent to $f(R)$ modified gravity theory, where $f(R)$ is a general function of the Ricci scalar \cite{Sotiriou:2008rp, DeFelice:2010aj}.  In addition, by adopting a conformal transformation it becomes equivalent to Einstein's theory of gravity minimally coupled to a scalar field with a canonical kinetic term and a specific potential given in terms of the function $f(R)$. This raises the question whether this scalar field, with gravitational origin, can act as an inflaton field. In this case, the arbitrary $f(R)$ may be constrained by the viability of inflationary potentials. Indeed, imposing the slow rolling conditions on $V(\phi)$ strongly limits $f(R)$ models.

In this paper we revisit the conditions that any $f(R)$ must fulfill in order to be a viable framework for early universe inflation. We analyze inflation in both the original and Einstein frames emphasizing that the scalar field picture is essential to study the detailed consequences of inflation like number of e-folds, spectral index, tensor-to-scalar ratio, and reheating after inflation, while in the $f(R)$ picture, one can only check the possibility that $f(R)$ may imply positively accelerating universe. We propose a new logarithmic $f(R)$ model satisfying the above mentioned conditions and discuss its consequences for inflation and reheating. Examples of logarithmic models were proposed inspired by quantum considerations, like the one loop-corrected effective action in the semiclassical approach to quantum gravity \cite{Alavirad:2013paa} or the renormalization group improvements of the Hilbert-Einstein action \cite{Frolov:2011ys} and others having various classical and cosmological motivations (check references in section \ref{sec:ourmodel}). Nevertheless, these models either do not comply with the conditions we set to satisfy or produce results incompatible with observation. The model proposed in this paper has an inflationary behavior very similar to that of Starobinsky's famous model \cite{Starobinsky:1980te} but with the added advantage of having no Ricci scalar singularity; in fact, when the model is expanded in a Taylor series around $R=0$ we find that the linear and quadratic terms are (almost) identical to those of Starobinsky, with the higher degree terms responsible for the mentioned singularity evasion.

The paper is organized as follows. In section \ref{sec:constructf} we discuss the conditions that an $f(R)$ must satisfy to account for inflation. We also study the equivalence of $f(R)$ gravity and general relativity coupled to a real scalar field.  Section \ref{sec:ourmodel} is devoted to analyzing inflation in our proposed $f(R)$. We show that this model leads, in the scalar field picture, to an inflationary potential with two mass parameters that accommodates the observational results. In section \ref{sec:reheating} we discuss the reheating process after inflation where only minimal coupling between gravity and matter fields is used to generate interaction between the inflaton and a scalar boson. Finally we give our conclusions in section \ref{sec:conclusions}. Throughout the paper we have adopted formulae for known inflationary parameters as they appeared in \cite{Planck:2013jfk, Ade:2015lrj}.

%%%%%%%%%%%%%%%%%%%%%%%%%%%%%%%%%%%%%%%%%%%%%%%%
\section{Constructing an inflationary \boldmath $f(R)$}\label{sec:constructf}
%%%%%%%%%%%%%%%%%%%%%%%%%%%%%%%%%%%%%%%%%%%%%%%%

The action for $f(R)$ gravity is given by
\begin{equation} \label{eq:faction}
S=\frac{1}{16\pi}\int d^4x\sqrt{-g}f(R)+\int d^4x\mathcal{L}_M(g_{\mu\nu},\chi),
\end{equation}
where $g$ is the determinant of the metric $g_{\mu\nu}$, $\mathcal{L}_M$ is the matter Lagrangian and $\chi$ represents matter fields. We will adopt the $(-,+,+,+)$ signature of the metric and the Planck system of units where $\hbar=c=G=1$. The corresponding field equations are given by
\begin{equation} \label{eq:feq}
f'R_{\mu\nu}-\frac{1}{2}fg_{\mu\nu}+\left(g_{\mu\nu}\square -\nabla_{\mu}\nabla_{\nu}\right)f'=8\pi T_{\mu\nu},
\end{equation}
where $f'=df(R)/dR$. Non-vanishing $\square f'$ indicates a propagating scalar degree of freedom whose dynamics is given by the trace of the above field equations
\begin{equation}\label{eq:trace}
3\square f'+f'R-2f=8\pi T,
\end{equation}
where $T$ is the trace of the stress-energy tensor $T_{\mu\nu}$.

\subsection{Equivalence with scalar-tensor theories}

As mentioned above, in addition to the usual 2 degrees of freedom of the metric in GR, the $f(R)$ modified gravity contains an extra degree of freedom which becomes manifest when a conformal transformation decouples it from the metric as a scalar field (giving rise to a spin-0 particle). This transformation relates $f(R)$ gravity to a conformally transformed Einstein term minimally coupled to a scalar field in the canonical form. This can be seen as follows: we start by rewriting the gravitational part of the action in the form:
\begin{equation}
S_G=\frac{1}{16\pi}\int d^4x\sqrt{-g} \left[ f(A) + f'(A)(R-A) \right],
\end{equation}
with the new field $A$. Variation with respect to $A$ gives $f''(A)(R-A)=0$ and if we require $f''(A)\neq 0$ then $A=R$ and action \eqref{eq:faction} is reproduced.

Defining $\sigma$ as $\sigma=f'(A)$ and assuming the invertibility of this relation, that is, $A$ could be expressed as a function of the field $\sigma$, then one gets the action in the \lq\lq Jordan frame"
\begin{equation} \label{eq:bdaction}
S_G=\frac{1}{16\pi}\int d^4x\sqrt{-g} \left[ \sigma R-U(\sigma) \right],
\end{equation}
where $U(\sigma)=A(\sigma)\sigma-f(A(\sigma))$. Equation \eqref{eq:bdaction} is the action of a Brans-Dicke theory with Brans-Dicke parameter equal to zero.

Now, to arrive at the Einstein frame, a conformal transformation of the metric is performed $g_{\mu\nu}\rightarrow \tilde{g}_{\mu\nu}= \sigma g_{\mu\nu}$. Such transformation implies a requirement that $\sigma \equiv f'(R) > 0$. Redefining the scalar field once more 
\begin{equation} \label{eq:fdef}
\sigma \equiv f'(R) = \textup{e}^{\sqrt{16\pi/3}\phi}
\end{equation}
and defining a potential function
\begin{equation} \label{eq:rpotential}
V(\phi)=\frac{Rf'-f}{16\pi f'^2},
\end{equation}
we arrive at the action in the Einstein frame
\begin{equation} \label{eq:einframe}
S_G=\frac{1}{16\pi}\int d^4x\sqrt{-\tilde{g}}\tilde{R}+\int d^4x \sqrt{-\tilde{g}}\left( -\frac{1}{2}\tilde{g}^{\mu\nu}\partial_{\mu}\phi\partial_{\nu}\phi-V(\phi) \right),
\end{equation}
in which the extra degree of freedom is represented as a scalar field in the canonical form minimally coupled to Einstein's gravity through metric $\tilde{g}_{\mu\nu}$. This was possible only when $f'(R)>0$, $f''(R)\neq0$ and when \eqref{eq:fdef} is invertible.

\subsection{Selection rules for $f(R)$}\label{subsec:condsonf}

We require a gravitational theory to
\begin{enumerate}[(i)]
	\item\label{i} Reduce to GR in its domain of validity, and hence to the Newtonian limit for the weak field approximation.
	\item\label{ii} Have the correct cosmological dynamics allowing a period of inflation in the early universe followed by a radiation domination era. This requirement is the main purpose of this paper.
	\item\label{iii} Be free of singularities.
\end{enumerate}

We shall use these general requirements to guide us build a model. An $f(R)$ could be cast in the form 
\begin{equation}\label{eq:F}
f(R)=R+F(R)
\end{equation}
such that $F(R)$ is  a certain correction to the Hilbert-Einstein action. Requirement (\ref{i}) implies $f(R\to0)\approx R+ {\cal O}(R^2) +\cdots$ so that there is no cosmological constant, the coefficient of the linear term is exactly 1 and the coefficients of the higher powers suppress them sufficiently within the experimental bounds. This also satisfies the condition
\begin{equation}\label{eq:minksol}
f(0)=0
\end{equation}
which is desirable to have a Minkowski space vacuum solution. As shown below, this leads to having the minimum of the potential equal to zero and thus having a Minksowski space solution in vacuum for the Einstein frame as well.

We can see what corresponds to condition \eqref{eq:minksol} in the scalar field picture. At the critical points $\phi_0$ of the potential, we have $V'_0 \equiv V'(\phi_0)=0$. By \eqref{eq:rpotential} one gets
\begin{equation}\label{eq:1atmin}
R_0=\frac{2f_0}{f'_0}=32\pi f'_0 V_0
\end{equation}
where the subscript zero always signifies the value at $\phi_0$. For $\phi_0$ to be a minimum of the potential we must have $V''_0>0$, that is, we can define a positive mass for the field as $M^2\equiv V''_0$, which reads
\begin{equation}\label{eq:rmfield}
M^2=\frac{1}{3}\left(\frac{1}{f''_0}-\frac{R_0}{f'_0}\right)>0.
\end{equation}
Equations \eqref{eq:1atmin} and \eqref{eq:rmfield} constitute the conditions for the minimum. If condition \eqref{eq:minksol} is satisfied, then we have $M^2=1/(3f''_0)>0$.

Since we aim to use the scalar field $\phi$ as an inflaton, requirement (\ref{ii}) in the scalar field picture brings in all the known conditions for single field inflationary potentials which include having a minimum for the potential at which the field can oscillate decaying into matter in the reheating process. Moreover, we will require that $f'(0)=1$ or $\phi_0=0$ to keep \emph{canonical} kinetic terms for the standard model particles that are naturally coupled to $\phi$ as we will show in section \ref{sec:reheating}.

Regarding requirement (\ref{iii}), Ricci scalar singularities can be studied easily in the Einstein frame. From eq. \eqref{eq:fdef} we see that $\phi$ can be viewed as a function of $R$ and this function is one-to-one since invertibility is assumed; hence, the dynamics of $\phi$ governed by potential \eqref{eq:rpotential} is the dynamics for $R$. Requiring the existence of an infinite potential barrier preventing $|R|$ from reaching infinity (i.e., $V(\phi(R\to\pm\infty))\to\infty$) is thus a way to keep the model free from Ricci scalar singularities. This condition translates to
\begin{equation}\label{eq:nosingularity}
\lim_{R\to\pm\infty}\frac{Rf'-f}{f'^2}=\infty.
\end{equation}

For models that are concerned with describing late time acceleration of the universe, reference \cite{Frolov:2008uf} should be consulted for a detailed analysis of curvature singularities of infrared corrected $f(R)$ gravity theories.

\subsection{Constraints from inflation}\label{subsec:conflation}

We attempt to attribute inflation wholly to gravitational effects, and so, we will assume that the contribution of matter is negligible ($T_{\mu\nu}=0$). The field equations \eqref{eq:feq} for an empty, flat Friedmann-Lema\^itre-Robertson-Walker (FLRW) universe give us the (modified) Friedmann equation
\begin{equation}\label{eq:frdmn}
\frac{1}{2}(Rf'-f)-3f'H^2-3H\dot{R}f''=0
\end{equation}
where $H\equiv\dot{a}/a$ is the Hubble parameter and $a(t)$ is the scale factor. The definition of $H$ gives us the identity $\ddot{a}/a=(1-\epsilon)H^2$ where $\epsilon$ is \lq\lq slow roll" parameter defined as\footnote{This terminology is actually borrowed from inflation by scalar field as shall be seen in subsection \ref{subsec:infleinstein}.}
\begin{equation}\label{eq:hslowroll}
\epsilon\equiv-\frac{\dot{H}}{H^2}.
\end{equation}
Inflation takes place when $\epsilon\ll1$ so that the acceleration of the scale factor is positive and $H$ is varying slowly giving an almost exponential expansion. Deviation from exact exponential expansion is measured by a hierarchy of parameters called the \emph{Hubble flow functions} (HFF) \cite{Schwarz:2001vv, Schwarz:2004tz} defined by $\epsilon_1=\epsilon$, $\epsilon_{i+1}=\dot{\epsilon_i}/(H\epsilon_i)$ where $i\geq1$. Having those parameters much less that unity ensures the mostly exponential growth of $a(t)$. Requiring the first two HFF's to be small suffices for simplifying \eqref{eq:frdmn} to a convenient form
\begin{equation}\label{eq:approxf}
\dot{H}\approx\frac{Rf'-2f}{24Rf''}
\end{equation}
where we have used that, for a flat FLRW metric, $R=6(2H^2+\dot{H})\approx12H^2$ and $\dot{R}\approx24H\dot{H}$. Expression \eqref{eq:approxf} requires $R\neq0$ and $f''\neq0$. From \eqref{eq:hslowroll} and \eqref{eq:approxf} we see that for inflation to occur, we must have
\begin{equation}\label{eq:infineq}
\epsilon\approx\frac{2f-Rf'}{2R^2f''}\ll1
\end{equation}
and inflation ends whenever \eqref{eq:infineq} turns into an equality. A viable model must have solutions for $\epsilon\approx1$, otherwise they will predict unending inflation ($\epsilon<1,\forall R$) or no inflation at all ($\epsilon>1,\forall R$). In case there are two solutions for $\epsilon\approx1$ then the lowest one should mark the end of inflation, we will call such solution $R_e$.

This analysis gives us general conditions on $f(R)$ theories aiming to describe inflation. The value of $H_e$ ($R_e$) should be contrasted with the observational bound \cite{Ade:2015lrj} that during inflation $H<7.18\times10^{-6}$ ($R<6.19\times10^{-10}$).

%%%%%%%%%%%%%%%%%%%%%%%%%
\section{A new logarithmic model}\label{sec:ourmodel}
%%%%%%%%%%%%%%%%%%%%%%%%%%

In accordance with the conditions discussed in section \ref{sec:constructf}, which guarantee the stability conditions analyzed in \cite{Pogosian:2007sw}, we propose the following $f(R)$ model
\begin{equation}\label{eq:mm}
f(R)=R+\frac{m^4}{3M^2}\left[\frac{R}{m^2}-\ln\left(1+\frac{R}{m^2}\right)\right]
\end{equation}
where $m$ and $M$ are free parameters. As will be seen, $m$ defines the scale below which inflation starts and will be determined from comparison with observation of various inflationary quantities. Parameter $M$ is the mass of the field (the inflaton) in the Einstein frame and will define the scale at which inflation ends.

It is notable the similarity between model \eqref{eq:mm} and the model presented in \cite{Miranda:2009rs} which was motivated by a generalization of the models proposed in \cite{Hu:2007nk} and \cite{Starobinsky:2007hu}. However, in contrast to model \eqref{eq:mm}, it does not produce the desired inflationary scenario. Other examples of logarithmic $f(R)$ models can be found in refs. \cite{Meng:2003en, Navarro:2005ux, Iorio:2007ee, Appleby:2008tv, Girones:2009nc, Hashemi:2010ut, Frolov:2011ys,Ivanov:2011np, Guo:2013swa, Guo:2013lka, Alavirad:2013paa, Astashenok:2013vza, Ben-Dayan:2014isa, Ketov:2014jta, Rinaldi:2014gha, Broy:2014xwa} and for examples of $f(R)$ inflationary models, check \cite{Starobinsky:1980te, Sotiriou:2008rp, DeFelice:2010aj, Capozziello:1993xn, Faulkner:2006ub, Cognola:2007zu, Nojiri:2010wj, Elizalde:2010ts, Hwang:2011kg, Ivanov:2011np, Motohashi:2012tt, Bamba:2013wj, Bamba:2012qi, Myrzakulov:2013hca, Sebastiani:2013eqa, Ben-Dayan:2014isa, Ketov:2014jta, Rinaldi:2014gha, Ellis:2014cma, Artymowski:2014gea, Capozziello:2014hia, Artymowski:2014nva, Broy:2014xwa, Sebastiani:2015kfa} and the references therein.

\subsection{Inflation in the $f(R)$ picture}\label{subsec:inflmetric}

\begin{figure}[t]
\centering
\psfig{file=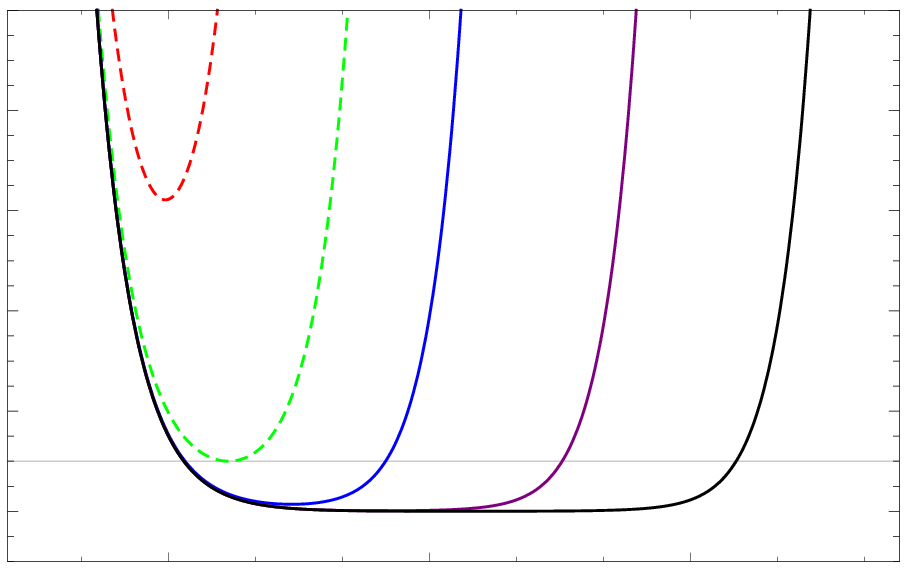, width=10cm}
\put(-8,7){{\scriptsize $10^{-3}$}}
\put(-83,7){{\scriptsize $10^{-6}$}}
\put(-158,7){{\scriptsize $10^{-9}$}}
\put(-233,7){{\scriptsize $10^{-12}$}}
\put(-165,-7){$R~(m_P^2)$}
\put(-290,32){{\scriptsize $0$}}
\put(-290,64){{\scriptsize $2$}}
\put(-290,96){{\scriptsize $4$}}
\put(-290,128){{\scriptsize $6$}}
\put(-290,160){{\scriptsize $8$}}
\put(-293,192){{\scriptsize $10$}}
\put(-307,96){\rotatebox{90}{$\epsilon$}}
\caption{Plot of $\epsilon$ against $R$ (in units of $m_P^2$) for different values of $m/M$ ($1,~\sim2.697,~10,10^2,10^3$) depicting the transition from no inflation (dashed) to inflation (solid) and the scale of $R$ at the start and end of inflation. $M$ is set equal to $10^{-6}$.}
\label{fig:epsilon_m}
\end{figure}

Following subsection \ref{subsec:conflation}, we find
\begin{equation}
\epsilon=\frac{(1+R/m^2)^2}{2R/m^2}\left[ 1+\frac{3M^2}{m^2}+\frac{1}{1+R/m^2}-\frac{2}{R/m^2}\ln(1+R/m^2) \right].
\end{equation}
One can see that there exists a certain upper limit for the ratio $M^2/m^2$ needed to have a region where $\epsilon<1$ and this limit is
\begin{equation}\label{eq:condrat}
\frac{M^2}{m^2}<g(R)
\end{equation}
where $g(R)$ is defined as
\begin{equation}
g(R)\equiv\frac{1}{3}\left( \frac{2R/m^2}{(1+R/m^2)^2}+\frac{2}{R/m^2}\ln(1+R/m^2)-\frac{1}{1+R/m^2}-1 \right).
\end{equation}
Studying the behavior of $g(R)$, it is found to have a global maximum $g(R)\lsim0.137$ and so a bound on $m$ is found to be $m>2.697~M$ as illustrated in figure \ref{fig:epsilon_m}.

Remarkably, the start ($R_i$) and end ($R_e$) of inflation are around the orders of magnitude of $m^2$ and $M^2$ respectively. Indeed, for $R\gg m^2$ one gets $\epsilon=\frac{R}{2m^2}(1+3M^2/m^2)>1$ and a solution ($R_i$) exists around $R\approx m^2$. The other solution ($R_e$) could be investigated as follows: for $R\ll m^2$ we get $\epsilon\approx3M^2/(2R)$, so inflation ends for $R<3M^2/2$, and the solution exists in the vicinity of $3M^2/2$; this estimation gets better whenever $m$ is set much larger than $M$. Precise determination of the values of the parameters $m$ and $M$ based on observational data will be detailed in the next section.

\subsection{Inflation in the Einstein frame}\label{subsec:infleinstein}

As advocated in the introduction, the Einstein frame has the benefit of studying inflation on the more familiar grounds of scalar field theory and provides the ability to compare with other scalar field inflationary models (e.g. those considered in \citep{Planck:2013jfk, Ade:2015lrj}). The Friedmann equation in the Einstein frame is
\begin{equation}\label{eq:vfriedmann}
\tilde{H}^2=\frac{8\pi}{3}\left[\frac{1}{2}\dot{\phi}^2+V(\phi)\right]
\end{equation}
and the equation of motion for the homogeneous \lq\lq inflaton" field is
\begin{equation}\label{eq:fieldmotion}
\ddot{\phi}+3\tilde{H}\dot{\phi}+V'(\phi)=0.
\end{equation}
Inflation takes place as long as the field is \lq\lq slowly rolling" on the potential such that\footnote{These conditions ensure that the field has the appropriate behavior for density and (negative) pressure to give the accelerated expansion known as inflation. This translates into a dominating friction term in \eqref{eq:fieldmotion}, hence slow rolling.}
\begin{equation}\label{eq:friction}
3\tilde{H}\dot{\phi}\approx-V'(\phi)
\end{equation}
and
\begin{equation}\label{vhubble}
\tilde{H}^2\approx\frac{8\pi}{3}V(\phi).
\end{equation}
The slow roll conditions, embodied in \eqref{eq:friction} and \eqref{vhubble}, could be cast into a form depending only on the potential of the field as \textit{slow roll parameters} having values less than unity during the inflationary period. These parameters are defined as
\begin{equation}\label{eq:vslowroll}
\begin{aligned}
\epsilon_{_V}\equiv\frac{1}{16\pi}\left(\frac{V'(\phi)}{V(\phi)}\right)^2\ll1,\qquad\eta_{_V}\equiv\frac{1}{8\pi}\frac{V''(\phi)}{V(\phi)}\ll1
\end{aligned}
\end{equation}
and inflation ends whenever $\epsilon_{_V}(\phi_e)\approx1$ or $\eta_{_V}(\phi_e)\approx1$ from which the value of the field at the end of inflation $\phi_e$ is calculated.

The purpose of postulating a period of rapid expansion of the early universe is to generate enough entropy consistent with the standard model of cosmology (hot Big Bang). This constraint translates into a number of \lq\lq e-folds" ($N\equiv\ln(a_e/a_*)$) the cosmic expansion has to cover from the instant of horizon crossing (signified by an asterisk subscript) before the end of inflation (signified by subscript $e$) and is estimated to be in the range 50--60. In terms of the potential, $N$ is expressed as
\begin{equation}\label{eq:nef}
N\approx-8\pi\int_{\phi_*}^{\phi_e}\frac{V(\phi)}{V'(\phi)}d\phi
\end{equation}
where $\phi_*$ is the value of the field at the horizon crossing and is estimated using this equation and $50<N<60$. With $\phi_*$ known, comparison with observation can now be made. The main observed quantities are:
\begin{equation}
\begin{aligned}
n_s\approx1+2\eta_{_V}(\phi_*)-6\epsilon_{_V}(\phi_*),\quad r\approx16\epsilon_{_V}(\phi_*),\quad \textup{and} \quad A_s\approx \frac{8}{3}\frac{V(\phi_*)}{\epsilon_{_V}(\phi_*)}
\end{aligned}
\end{equation}
where $n_s$ is the \emph{scalar spectral index}, $r$ is the \emph{tensor-to-scalar} ratio and $A_s$ is the amplitude of the power spectrum of curvature perturbations.

\begin{figure}[t]
\centering
\psfig{file=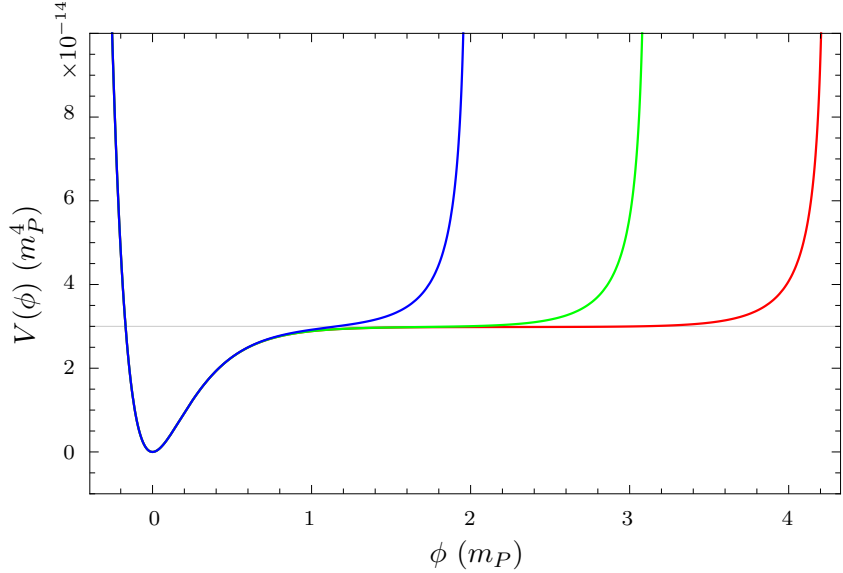, width=10cm}
\put(-23,7){{\scriptsize $4$}}
\put(-83,7){{\scriptsize $3$}}
\put(-142,7){{\scriptsize $2$}}
\put(-202,7){{\scriptsize $1$}}
\put(-260,7){{\scriptsize $0$}}
\put(-155,-7){$\phi~(m_P)$}
\put(-292,32){{\scriptsize $0$}}
\put(-292,64){{\scriptsize $2$}}
\put(-292,96){{\scriptsize $4$}}
\put(-292,128){{\scriptsize $6$}}
\put(-292,160){{\scriptsize $8$}}
\put(-295,175){\rotatebox{90}{{\footnotesize $\times10^{-14}$}}}
\put(-312,75){\rotatebox{90}{$V(\phi)~(m_P^4)$}}
\caption{$V(\phi)$ (in units of $m_P^4$) against $\phi$ (in units of $m_P$) for different values of $m$: $10^{-4}$, $10^{-3}$ and $10^{-2}$ respectively from left to right. $M$ is set equal to $10^{-6}$.}
\label{fig:p_m}
\end{figure}

Using \eqref{eq:fdef} and \eqref{eq:rpotential} we find the potential corresponding to \eqref{eq:mm}
\begin{equation}\label{eq:vpotential}
V(\phi)=\frac{m^2}{16\pi}\textup{e}^{-2\sqrt{16\pi/3}\phi}\left\{ 1-\textup{e}^{\sqrt{16\pi/3}\phi}-\frac{m^2}{3M^2} \ln \left[ 1+\frac{3M^2}{m^2}\left(1-\textup{e}^{\sqrt{16\pi/3}\phi} \right) \right] \right\}.
\end{equation}
Figure \ref{fig:p_m} depicts the shape of the potential for different values of $m$ at fixed $M$. We can see that the potential flattens at about $V\approx3\times10^{-14}$ starting around $\phi\approx1$. The smaller $m$ is, the shorter this \lq\lq inflationary plateau" becomes, which is similar to the dependence of inflation on $m$ in the $f(R)$ picture discussed in subsection \ref{subsec:inflmetric}. We also see that the potential goes to infinity at $\phi(R\to\infty)$ as required in subsection \ref{subsec:condsonf}.

As can be seen from equation \eqref{eq:vpotential}, other than the factor of $m^2$, the potential depends on the ratio $m/M$. Therefore, $\epsilon_{_V}$, $\eta_{_V}$ and $N$ depend only on this ratio that can be treated as a single parameter. Only $A_s$ depends on both $m$ and $m/M$. 

\begin{figure}[t]
\centering
\psfig{file=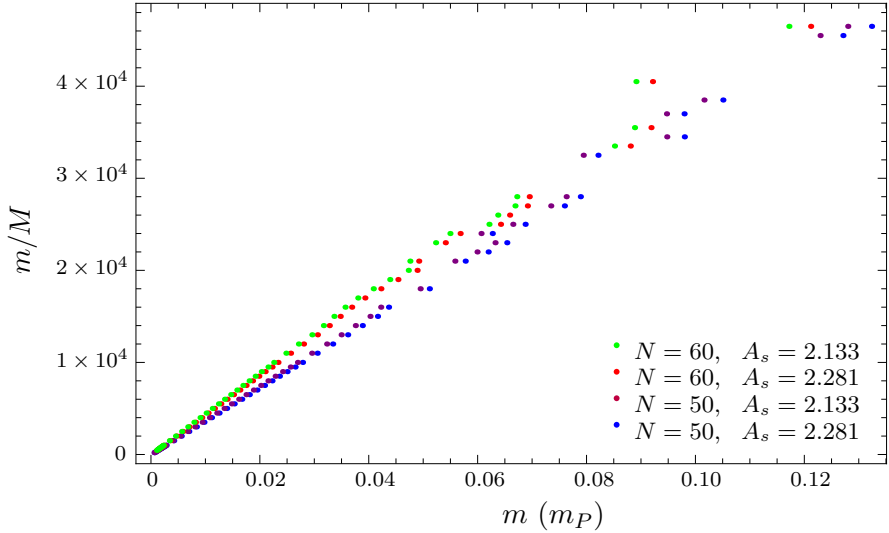, width=10cm}
\put(-37,7){{\scriptsize $0.12$}}
\put(-78,7){{\scriptsize $0.10$}}
\put(-118,7){{\scriptsize $0.08$}}
\put(-159,7){{\scriptsize $0.06$}}
\put(-200,7){{\scriptsize $0.04$}}
\put(-241,7){{\scriptsize $0.02$}}
\put(-278,7){{\scriptsize $0$}}
\put(-145,-7){$m~(m_P)$}
\put(-290,17){{\scriptsize $0$}}
\put(-312,52){{\scriptsize $1\times10^4$}}
\put(-312,87){{\scriptsize $2\times10^4$}}
\put(-312,122){{\scriptsize $3\times10^4$}}
\put(-312,157){{\scriptsize $4\times10^4$}}
\put(-330,87){\rotatebox{90}{$m/M$}}
\put(-105,55){{\huge {\color{green} $\cdot$}} {\footnotesize $N=60,~~A_s=2.133$}}
\put(-105,45){{\huge {\color{red} $\cdot$}} {\footnotesize $N=60,~~A_s=2.281$}}
\put(-105,35){{\huge {\color{purple} $\cdot$}} {\footnotesize $N=50,~~A_s=2.133$}}
\put(-105,25){{\huge {\color{blue} $\cdot$}} {\footnotesize $N=50,~~A_s=2.281$}}
\caption{The ratio $m/M$ vs $m$ (in units of $m_P$) satisfying the observational constraints on $N$, $n_s$, $r$ and $A_s$. The four sets of points correspond to: $N=50$ (lower pair) and $N=60$ (upper pair). Each pair corresponds to the two observational limits of $A_s$.}
\label{fig:mandratio}
\end{figure}

\begin{figure}[t]
\centering
\psfig{file=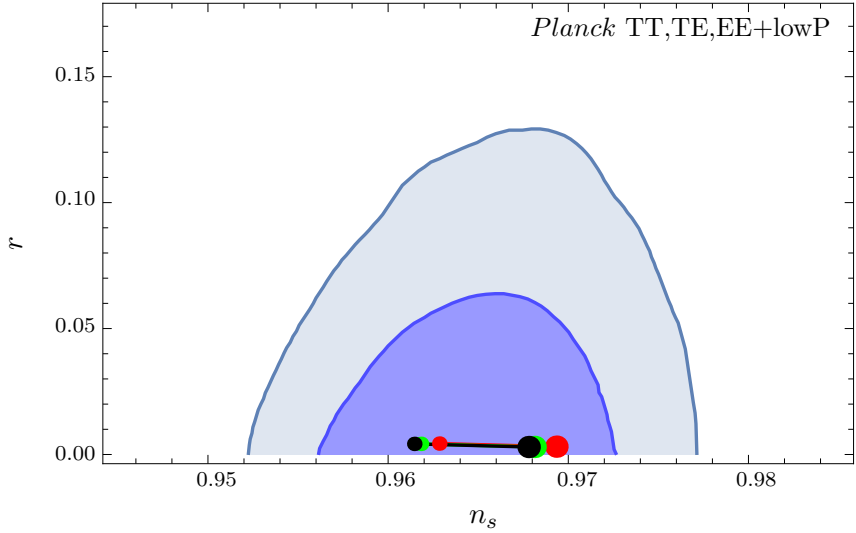, width=10cm}
\put(-122,177){{\small $Planck~\textup{TT,TE,EE+lowP}$}}
\put(-45,7){{\scriptsize $0.98$}}
\put(-112,7){{\scriptsize $0.97$}}
\put(-180,7){{\scriptsize $0.96$}}
\put(-247,7){{\scriptsize $0.95$}}
\put(-145,-7){$n_s$}
\put(-300,17){{\scriptsize $0.00$}}
\put(-300,65){{\scriptsize $0.05$}}
\put(-300,113){{\scriptsize $0.10$}}
\put(-300,160){{\scriptsize $0.15$}}
\put(-317,96){\rotatebox{90}{$r$}}
\caption{A plot of the tensor-to-scalar ratio $r$ versus the scalar spectral index $n_s$ obtained from the model for $m/M\sim10^{2}$ (red), $\sim10^{3}$ (green) and $\sim10^{4}$ (black), where the small and big dots are evaluated at $N=50$ and $N=60$ respectively, compared with experimental data from the \textit{Planck} mission, 2015 (blue background).}
\label{fig:data}
\end{figure}

In figure \ref{fig:mandratio}, we plot $m/M$ versus $m$ satisfying the observational constraints: $50\leq N\leq 60$, $r<0.1$, $n_s = 0.9645\pm0.0049$ and $A_s=(2.207\pm0.074)\times10^{-9}$ \cite{Ade:2015lrj}\footnote{We use the {\it Planck} TT,TE,EE+lowP vlaues in particular.}. For a given $N$ and $A_s$, the plot is (almost) a straight line, which shows that the inflaton mass $M$ must be a constant measured by the inverse of the slope. We find that $2.35\times10^{-6}<M<2.76\times10^{-6}$, or just $M\sim10^{-6}$. Futhermore, the range of allowed values of $m/M$ is roughly $10^2\lsim m/M\lsim10^4$, which implies that $10^{-4} \lsim m \lsim 10^{-2}$. Figure \ref{fig:data} shows three representative results of the $r$ vs $n_s$ corresponding to the allowed range of values of $m/M$ in comparison with the observational data from \citep{Ade:2015lrj}. We see that $0.9615<n_s<0.9693$ and $r\sim10^{-3}$. Moreover, calculating $\phi_*$ and $\phi_e$ for the model, one finds that $\phi_*\approx1.1~m_P$ and $\phi_e\approx0.2~m_P$, i.e. the horizon crossing occures when the inflaton field is around the Planck scale and inflation ends about one order of magnitude less than that.
%----------------------------------------------
%-------------------------------------------------

%%%%%%%%%%%%%%%%%%%
\section{Reheating after inflation}\label{sec:reheating}
%%%%%%%%%%%%%%%%%%%

We now consider the reheating  process after inflation. As inflation ends, the inflaton field smoothly enters an era of damped oscillation around the Minkowski minimum, eventually decaying and reheating the universe. The method described in this section is based on minimal coupling of gravity with matter fields.

To study the decay of the inflaton, we must consider the second part of the total action given in \eqref{eq:faction}, where the Lagrangian  of some matter field, e.g., scalar boson $\chi$, is given by
\begin{equation}
\mathcal{L}_M(g_{\mu\nu},\chi) = - \frac{1}{2} g_{\mu \nu}  \partial^\mu \chi \partial^\nu \chi - V(\chi)
\end{equation}
where $V(\phi)$ is the potential for matter. Here we assume minimal interaction between gravity and matter fields. In the Einstein frame the above Lagrangian takes the form
\begin{eqnarray}
\tilde{\mathcal{L}}_M &=& - \frac{1}{2} \tilde{g}_{\mu \nu}\textup{e}^{-\sqrt{16 \pi/3}\phi/m_P} \partial^\mu \chi \partial^\nu \chi - V(\chi) \nonumber\\
&=& - \frac{1}{2} \tilde{g}_{\mu \nu} \left[1- \frac{\sqrt{16\pi/3}}{m_P} \phi + \cdots \right]~  \partial^\mu \chi \partial^\nu \chi - V(\chi),
\end{eqnarray}
where we have restored the Planck mass ($m_P=1/\sqrt{G}$) for clarity. The following coupling between the inflaton and $\chi$ scalar boson is generated
\begin{equation}
\lambda_{\phi \chi\chi} =  \frac{\sqrt{4\pi/3}}{m_P} p_{_{1}\mu} p_{_2}^\mu,
\end{equation}
where $p_{_1}$ and $p_{_2}$ are the 4-momenta of $\chi$ particles. Such a general mechanism of generating coupling between the inflaton and a scalar boson is independent from any specific $f(R)$ model save for the conditions $M^2>0$ (having a minimum for the potential) and $f'(R_0)=1$ (the minimum is at $\phi=0$).
%----------------------------------------------
\begin{figure}[t]
\centering
\psfig{file=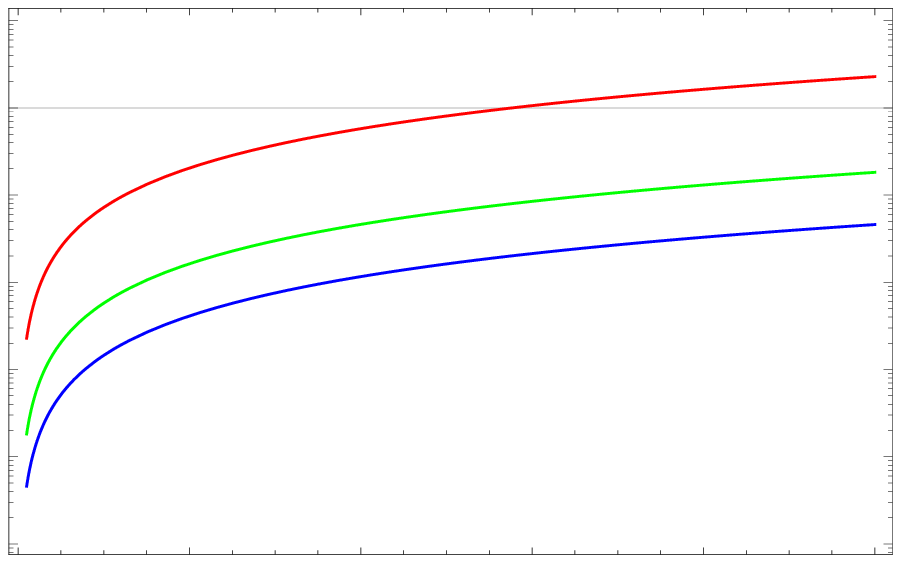, width=10cm}
\put(-13,7){{\scriptsize $10$}}
\put(-64,7){{\scriptsize $8$}}
\put(-119,7){{\scriptsize $6$}}
\put(-173,7){{\scriptsize $4$}}
\put(-227,7){{\scriptsize $2$}}
\put(-280,7){{\scriptsize $0$}}
\put(-167,-7){$M~(\textup{GeV})$}
\put(-297,21){{\scriptsize $10^{4}$}}
\put(-297,49){{\scriptsize $10^{5}$}}
\put(-297,76){{\scriptsize $10^{6}$}}
\put(-297,101){{\scriptsize $10^{7}$}}
\put(-297,129){{\scriptsize $10^{8}$}}
\put(-297,157){{\scriptsize $10^{9}$}}
\put(-300,185){{\scriptsize $10^{10}$}}
\put(-314,80){\rotatebox{90}{$T_R~(\textup{GeV})$}}
\caption{Reheating temperature as function of the inflaton mass for $m_\chi/M$ equal to 0.4, 0.1 and  0.01 respectively from top to bottom.}
\label{fig:TR}
\end{figure}
%-------------------------------------------------
The decay rate of $\phi \to \chi \chi$ is given by
\begin{equation}
\Gamma = \frac{1}{8 \pi} \frac{m_\chi^4}{M m_p^2} \sqrt{1- \left(\frac{2 m_\chi}{M}\right)},
\end{equation}
where $m_\chi$ is the mass of the scalar boson $\chi$. The reheating temperature from the decay of the inflaton is given by
\begin{equation}
T_R \approx \frac{(8 \pi)^{1/4}}{7} \sqrt{m_P \Gamma}.
\end{equation}
Since the inflaton mass is fixed $M \approx {\mathcal O}(10^{13})$ GeV, the mass of the scalar field $\chi$ should be constrained by $m_\chi \lsim M/2$. In Fig. \ref{fig:TR} we display the reheating temperature $T_R$ as function of the inflaton mass and we find that the reheating temperature does not exceed $10^9$ GeV which is consistent with the upper bound from gravitino production \citep{Khlopov:1984pf, Ellis:1984eq}. The lower bound on the reheating temperature from the Big Bang nucleosynthesis \cite{Sarkar:1995dd} being $\mathcal{O}(1)$ MeV sets a lower bound on the boson mass to be of order $10^7$ GeV.

\section{Conclusions}\label{sec:conclusions}

In this paper, we have proposed a model that quite accurately meets the observed constraints on inflation and describes the transition to a radiation domination era through a phase of reheating. After providing a detailed discussion for the constraints imposed on any viable inflationary $f(R)$, we considered a new logarithmic $f(R)$ model with two mass parameters and analyzed the associated inflation in both the original $f(R)$ picture and the Einstein frame. We showed that in our model the horizon crossing takes place at scale $\sim m_p$ and leads to results consistent with observations from \textit{Planck} mission 2015, namely $0.9615<n_s<0.9693$ and $r\sim10^{-3}$.  After inflation ends, the inflaton oscillates about the Minkowski vacuum decaying into a scalar boson with mass: $10^7~ \textup{GeV}<m_\chi<0.5~\times10^{13}~\textup{GeV}$, thereby reheating the universe with temperature: $1~\textup{MeV}<T_R<10^{9}~\textup{GeV}$ in accordance with bounds of Big Bang nucleosynthesis and gravitino production. We emphasized that in our model, a coupling between matter fields and inflaton can be naturally generated through minimal coupling to gravity.

\section*{Acknowledgements}
This work was partially supported by the STDF project 13858 and the ICTP grant AC-80. MA and MS would like to thank W. Abdallah, M. Ashry and A. Moursy for helpful discussions.
%%%%%%%%%%%%%%%%%%%%%%%%%%%%%%%%%%%%%%%%%%%%%%%%%%%%%%%%%%%%%%%%%

\bibliographystyle{JHEP}
	\bibliography{Bib}

\end{document}